\pdfoutput=1
\documentclass[%
 reprint,
 prd,
 superscriptaddress,
 preprintnumbers,
 bibnotes,
 amsmath,amssymb,
 aps,
]{revtex4-1}
\usepackage{graphicx}
\usepackage{color}
\usepackage{array}
\usepackage{slashed}
\usepackage{tikz}
\usepackage{hyperref}
\usepackage[caption=false]{subfig}
\usepackage{arydshln}
\usepackage{comment}
\usepackage{yhmath}

\usetikzlibrary{shapes.geometric,arrows,decorations.pathmorphing,decorations.markings,patterns,calc}

\usepackage{color}

\newif\ifdraft
\drafttrue

\def\spa#1.#2{\left\langle#1\,#2\right\rangle}
\def\spb#1.#2{\left[#1\,#2\right]}

\def\e{\epsilon}

\font\tenshuffle=shuffle10 \font\sevenshuffle=shuffle7 \font\fiveshuffle=shuffle7 at 5pt
\def\shuffle{{%
\def\Dshuffle{\mathbin{\hbox{\tenshuffle\char'001}}}%
\def\Sshuffle{\mathbin{\hbox{\sevenshuffle\char'001}}}%
\def\SSshuffle{\mathbin{\hbox{\fiveshuffle\char'001}}}%
\mathchoice{\Dshuffle}{\Dshuffle}{\Sshuffle}{\SSshuffle}}}

\newcommand{\eq}{\begin{equation}}
\newcommand{\eqe}{\end{equation}}
\newcommand{\eqa}{\begin{eqnarray}}
\newcommand{\eqae}{\end{eqnarray}}

\newcommand{\bea}{\begin{eqnarray}}
\newcommand{\eea}{\end{eqnarray}}

\newbox\charbox
\newbox\slabox
\def\s#1{{      
        \setbox\charbox=\hbox{$#1$}
        \setbox\slabox=\hbox{$/$}
        \dimen\charbox=\ht\slabox
        \advance\dimen\charbox by -\dp\slabox
        \advance\dimen\charbox by -\ht\charbox
        \advance\dimen\charbox by \dp\charbox
        \divide\dimen\charbox by 2
        \raise-\dimen\charbox\hbox to \wd\charbox{\hss/\hss}
        \llap{$#1$}
}}

\def\be{\begin{equation}}
\def\ee{\end{equation}}
\def\ba{\begin{eqnarray}}
\def\ea{\end{eqnarray}}
\def\nl{\nonumber\\}

\def\<{\langle}
\def\>{\rangle}

\def\bea#1\ea{\begin{eqnarray}#1\end{eqnarray}}
\def\be#1\ee{\begin{equation}#1\end{equation}}
\def\ba#1\ea{\begin{align}#1\end{align}}

\def\nl{\nonumber\\}

\def\yz#1\yz {{\color{blue} [[YZ: #1]] }}

\usepackage{cleveref}

\usepackage{orcidlink}

\begin{document}

\title{
New Factorizations  of Yang-Mills Amplitudes 
} 

\author{Alfredo Guevara \!\orcidlink{0000-0002-8963-6560}}

\email{aguevara@ias.edu}

\affiliation{
School of Natural Sciences, Institute for Advanced Study, Princeton NJ 08540, USA}

\author{Yong Zhang \!\orcidlink{0000-0002-3522-0885}}

\email{yzhang@perimeterinstitute.ca}

\affiliation{Perimeter Institute for Theoretical Physics, Waterloo, ON N2L 2Y5, Canada.}

\affiliation{
School of Physical Science and Technology, Ningbo University, Ningbo 315211, China.}

\begin{abstract}

We propose a new factorization pattern for tree-level Yang-Mills (YM) amplitudes, where they decompose into lower-point amplitudes by setting non-planar Mandelstam variables to zero. This approach manifests the hidden zeros of YM amplitudes recently identified, and takes the form of a novel soft theorem. Furthermore, by setting specific Lorentz products involving polarization vectors to zero, the amplitudes further reduce to a sum of products of three currents. These novel factorizations provide a fresh perspective on the structure of YM amplitudes, potentially enhancing our understanding and calculation of perturbative amplitudes.

\end{abstract}

\maketitle

\section*{Introduction }  

Scattering amplitudes in quantum field theory factorize into sums of products of lower-point amplitudes at the residues of their poles, reflecting the fundamental property of locality. This factorization enables recursive computation of amplitudes, making it a powerful method in modern scattering amplitude research. A notable application of this concept is the Britto-Cachazo-Feng-Witten  recursion \cite{Britto:2005fq}, which introduces an auxiliary variable in the complex plane to control amplitude factorization, leading to significant advancements.

At the loop level, (generalized) unitarity cuts are employed to determine the loop integrand of amplitudes \cite{Bern:1994zx,Bern:1994cg}. The primary difference from conventional factorization is the substitution of a loop propagator for a tree propagator on-shell. Thus, studying factorizations also aids in understanding unitarity cuts at the loop level. Recently, factorizations have been utilized to determine Mellin amplitudes in AdS space (cf. \cite{Goncalves:2014rfa, Alday:2023kfm, Cao:2023cwa}).  

Given the importance of factorizations in scattering amplitude studies, new types of factorizations have been investigated. In \cite{Cachazo:2021wsz}, it was proposed that by setting specific Mandelstam variables to zero, amplitudes for biadjoint $\phi^3$ theory, the Nonlinear Sigma Model (NLSM), and a special Galileon theory \cite{Cachazo:2014xea} split into three currents, a property termed semi-locality. This differs from conventional factorizations as no residues are computed \footnote{The smoothly splitting amplitudes for $\phi^p$ theories are discussed in \cite{GimenezUmbert:2024jjn}. See more recent progress in \cite{GimenezUmbert:2025ech,Early:2025ivr}}.

In \cite{Arkani-Hamed:2023swr}, another form of three-part splitting was discovered for $\text{Tr}(\phi^3)$ theory and NLSM. By setting certain two-point Mandelstam variables in a rectangle to zero and then turning back on one of them, the amplitudes split into three parts near their zeros. These two types of three-part splitting for scalar theories are unified by a so-called 2-split proposed in \cite{Cao:2024gln, Cao:2024qpp}, which asserts that amplitudes factorize into two currents under certain subspaces of the Mandelstam variables. Applying the 2-splits in different ways reproduces both types of three-part splitting found in \cite{Cachazo:2021wsz} and \cite{Arkani-Hamed:2023swr}. 
Additionally, significant progress has been made in extending such split factorizations to loop orders by analyzing the curve integral formulation for ${\rm Tr}(\phi^3)$ amplitudes, as discussed in \cite{Arkani-Hamed:2023swr, Arkani-Hamed:2023jry, Arkani-Hamed:2024nhp, Arkani-Hamed:2024yvu, Arkani-Hamed:2024fyd}.

Beyond scalar theories, hidden zeros in tree-level Yang-Mills (YM) 
amplitudes were also proposed in \cite{Arkani-Hamed:2023swr} by setting specific Mandelstam variables in a rectangle to zero along with their corresponding Lorentz products involving polarizations. Subsequently, the 2-split for YM and GR amplitudes was designed in \cite{Cao:2024gln, Cao:2024qpp} by turning on some vanishing Mandelstam variables while setting additional Lorentz products involving polarizations to zero. The hidden zeros in scalar theories have been further studied in \cite{Bartsch:2024amu,Li:2024qfp} using double copy relations \cite{Kawai:1985xq,Bern:2008qj} and uniqueness bootstrap methods \cite{Rodina:2024yfc}, which are also claimed to be applicable to the hidden zeros in amplitudes of  YM, GR, and other theories.

In this paper, we further investigate a new type of factorization of YM amplitudes by setting specific Mandelstam variables in a rectangle to zero, identical to the constraint used to produce zeros in certain scalar amplitudes \cite{Arkani-Hamed:2023swr}. We found that under this constraint, YM amplitudes always factorize into sums of gluings of lower-point YM amplitudes. In addition to the summation over all possible states of the polarization of the internal gluon that is exchanged, as in conventional factorization, our factorization further includes a summation over different gluon pairs appearing in the vanishing Mandelstam variable rectangle.

\begin{figure*}[!htbp]
\centering
\begin{tikzpicture}

\begin{scope}[xshift=-11cm]

 
    \draw[thick, decorate, decoration={snake, amplitude=.7mm, segment length=1.5mm}] (0,0) -- +(0,2) node [above] {$m{+}1$} ;
    
        \draw[thick, decorate, decoration={snake, amplitude=.7mm, segment length=1.5mm}] (0,0) -- +(0,-2) node [below] {$n$}  ;

        \draw[blue, thick, decorate, decoration={snake, amplitude=.7mm, segment length=1.5mm}] (0,0) -- (66:2) node [above right] {$m$}  ;
                
       \draw[blue, thick, decorate, decoration={snake, amplitude=.7mm, segment length=1.5mm}] (0,0) -- (-66:2) node [below right] {$1$}  ;

    \draw[blue, thick, decorate, decoration={snake, amplitude=.7mm, segment length=1.5mm}] (0,0) -- (42:2) node [above right] {$m{-}1$}  ;
                
       \draw[blue, thick, decorate, decoration={snake, amplitude=.7mm, segment length=1.5mm}] (0,0) -- (-42:2) node [below right] {$2$}  ;

              \filldraw[blue] (0:1.5) circle (1pt) ; 
              \filldraw[blue] (20:1.5) circle (1pt) ; 
                   \filldraw[blue] (-20:1.5) circle (1pt) ; 

   \draw[red, thick, decorate, decoration={snake, amplitude=.7mm, segment length=1.5mm}] (0,0) -- (114:2) node [above left] {$m{+}2$}  ;
                
  \draw[red, thick, decorate, decoration={snake, amplitude=.7mm, segment length=1.5mm}] (0,0) -- (-114:2) node [below left] {$n{-}1$}  ;

     \draw[red, thick, decorate, decoration={snake, amplitude=.7mm, segment length=1.5mm}] (0,0) -- (138:2) node [above left] {$m{+}3$}  ;
                
  \draw[red, thick, decorate, decoration={snake, amplitude=.7mm, segment length=1.5mm}] (0,0) -- (-138:2) node [below left] {$n{-}2$}  ;

              \filldraw[red] (180:1.5) circle (1pt) ; 
              \filldraw[red] (160:1.5) circle (1pt) ; 
                   \filldraw[red] (200:1.5) circle (1pt) ; 

  \node at ( 4,0) {$
  \xrightarrow[
   \substack{ \forall\, 
  {\color{blue} 1\leq a \leq m }
   \\  
  {\color{red} m{+}2 \leq b \leq n{-}1 } }
  ]{s_{ab}\to 0 }
\sum\limits_{ 
    \substack{ \forall\, 
  {\color{blue} 1\leq i \leq m }
   \\  
  {\color{red} m{+}2 \leq j \leq n{-}1 } 
 \\
 \rho\in S_{m{-}1}
 \\
 \epsilon_{\hat j}
  }
  }
  %
  $};

\filldraw[color=gray!90, fill=gray!45, very thick ](0,0) circle (1.1);
\node at (0,0) {$A_n$};
            
 \end{scope}

    \draw[thick, decorate, decoration={snake, amplitude=.7mm, segment length=1.5mm}] (0,0) -- (0,2) node [above] {$m{+}1$} ;
    
        \draw[thick, decorate, decoration={snake, amplitude=.7mm, segment length=1.5mm}] (0,0) -- (0,-2) node [below] {$n$}  ;

        \draw[blue, thick, decorate, decoration={snake, amplitude=.7mm, segment length=1.5mm}] (0,0) -- (12:2) node [right] {$\rho(i{+}1)$}  ;
                
       \draw[blue, thick, decorate, decoration={snake, amplitude=.7mm, segment length=1.5mm}] (0,0) -- (-12:2) node [right] {$\rho(i{-}1)$}  ;

          \draw[blue, thick, decorate, decoration={snake, amplitude=.7mm, segment length=1.5mm}] (0,0) -- (66:2) node [above right] {$\rho(m)$}  ;
                
       \draw[blue, thick, decorate, decoration={snake, amplitude=.7mm, segment length=1.5mm}] (0,0) -- (-66:2) node [below right] {$\rho(1)$}  ;
       
       \filldraw[blue] (39:1.5) circle (1pt) ; 
              \filldraw[blue] (51:1.5) circle (1pt) ; 
                     \filldraw[blue] (27:1.5) circle (1pt) ;

           \filldraw[blue] (-39:1.5) circle (1pt) ; 
              \filldraw[blue] (-51:1.5) circle (1pt) ; 
                     \filldraw[blue] (-27:1.5) circle (1pt) ;

        \draw[thick, decorate, decoration={snake, amplitude=.7mm, segment length=1.5mm}]  (-180:4)--(0,0)  ;     

       \draw[blue, thick, decorate, decoration={snake, amplitude=.7mm, segment length=1.5mm}] (-180:4)--+(120:1.5)  node [above left] {$i$}  ;
       \draw[red, thick, decorate, decoration={snake, amplitude=.7mm, segment length=1.5mm}] (-180:4)--+(-120:1.5)  node [below left] {$j$}  ;
        
    \node at ( -2.2,-.3) {$\hat j$};
      \node at (- 3.0,-.3) {$-\hat j$};

  \draw[red, thick, decorate, decoration={snake, amplitude=.7mm, segment length=1.5mm}] (0,0) -- (158:2) node [above left] {$j{-}1$}  ;
  
    \draw[red, thick, decorate, decoration={snake, amplitude=.7mm, segment length=1.5mm}] (0,0) -- (-158:2) node [below left] {$j{+}1$}  ;

     \draw[red, thick, decorate, decoration={snake, amplitude=.7mm, segment length=1.5mm}] (0,0) -- (112:2) node [above left] {$m{+}2$}  ;
     
          \draw[red, thick, decorate, decoration={snake, amplitude=.7mm, segment length=1.5mm}] (0,0) -- (-112:2) node [below left] {$n{-}1$}  ;

                     \filldraw[red] (135:1.5) circle (1pt) ; 
              \filldraw[red] (145:1.5) circle (1pt) ; 
                     \filldraw[red] (125:1.5) circle (1pt) ; 
                     
              \filldraw[red] (-135:1.5) circle (1pt) ; 
              \filldraw[red] (-145:1.5) circle (1pt) ; 
                     \filldraw[red] (-125:1.5) circle (1pt) ;

\filldraw[color=gray!90, fill=gray!45, very thick ](0,0) circle (.9);
\node at (0,0) {$A_{n{-}1}$};

\end{tikzpicture}
\caption{
\label{graph}
New factorization of YM amplitudes: An \(n\)-point Yang-Mills amplitude factorizes into a sum over contributions, each involving a 3-point amplitude and an \((n-1)\)-point amplitude. The sum runs over all permutations \(\rho\) of the \( (m-1) \) external gluons. The internal gluon \(\hat{j}\) is exchanged in two diagrams.  For simplicity, some rational functions of Mandelstam variables are omitted in this schematic representation.
\\
The first graph in this figure also provides an intuitive explanation for the constraints on the Mandelstam variables in \( h_m \), as defined in \eqref{eq:constrhm}. Selecting any two non-adjacent external gluon legs (e.g., \( m{+}1 \) and \( n \)) in a color-ordered amplitude naturally partitions the remaining legs into two sets. Setting all two-particle Mandelstam variables that involve legs from both sets to zero gives \( h_m = 0 \).
}
\end{figure*}

We refer to the contribution from each gluon pair as a {\it gluon pair contribution}. By eliminating all Lorentz products involving polarizations associated with the vanishing Mandelstam variables in the rectangle, we can observe how all gluon pair contributions vanish directly in our factorization formulas, thus making the hidden zeros manifest. Conversely, since the proposal of zeros in YM amplitudes in \cite{Arkani-Hamed:2023swr}, an important question has been the behavior of amplitudes near these zeros when Lorentz products involving polarizations are turned on back. Our factorization formulas address this question precisely, in fact, {the numerators of our expression naturally take the form of the famous Leading and Subleading Soft Gluon theorems \cite{Weinberg:1965nx,Cachazo:2014fwa,Casali:2014xpa}.}


Our factorization formulas for YM amplitudes differ from the 2-split proposed in \cite{Cao:2024gln, Cao:2024qpp} in that we only need to set certain Mandelstam variables to zero while allowing all Lorentz products involving polarization to remain. 
Additionally, our factorization formulas typically include a summation of gluing contributions instead of a single product of two amplitudes.
 However, combining our factorization formulas with the 2-split formulas yields another elegant result. By imposing conditions on Lorentz products such that only a single gluon pair contribution survives in our factorization summation for YM amplitudes, and then further eliminating additional Lorentz products involving polarizations, the entire YM amplitude factorizes into a sum of three currents. Continued study of our new factorization should uncover more properties of YM amplitudes, providing a novel description.

\section*{New Factorizations} 

Consider a matrix of Mandelstam variables $ s_{ab} := k_a \cdot k_b $, denoted as $ h_m $ with $ 1 \leq m \leq n-3 $,
\begin{align}
    h_m:=\begin{Bmatrix} 
    \hspace{-0.0cm} s_{1\,m+2}\, , \,s_{1\,m+3}\, , \ldots , s_{1\,n-1}\,\, , \\
    \hspace{-0.0cm} s_{2\,m+2}\, , \,s_{2\,m+3}\, , \ldots ,  s_{2\,n-1}\,\, , \\
    \hspace{-0.5cm}
    \vdots \hspace{1.1cm} \vdots \hspace{1.7cm} \vdots\\
    \hspace{-0.3cm}\hspace{0.3cm}s_{m\,m+2} , s_{m\,m+3} , \ldots , s_{m\,n-1}  ~~
    \end{Bmatrix}\,.
    \label{eq:constrhm}
\end{align}
As established in \cite{Arkani-Hamed:2023swr}, setting $ h_m = 0 $ \footnote{We assume the spacetime dimension is general enough to ignore the Gram determinant constraints.} results in the vanishing of $n$-point doubly color-ordered biadjoint $\phi^3$ amplitudes with two canonical orderings, as well as many other scalar amplitudes.  YM amplitudes do not vanish under this constraint; in this paper, we analyze how they factorize.

\subsection*{Cases $ m = 1, 2 $}

To introduce the factorizations we will first fully lay out the case $ m = 1$, and present the general formula in a later section.
We find
\ba 
&A({\mathbb I}_n)  \xrightarrow{s_{13}=s_{14}\ldots=s_{1\,n\!-\!1}=0}
\\ \nonumber 
& \quad \frac{(-1)^{n+1}}{s_{12}} \sum_{j=3}^{n-1}  \sum_{\e_{\hat j}} 
 A(1j\, -\!\hat j) A(23\ldots j\!-\!1\, \hat j \,j\!+\!1  \ldots n)\,,
\ea 
{where \( A(ij - \hat{j}) \) is the three-point amplitude. We will give a few explicit expressions of the sum over polarizations below and relate it to a soft gluon theorem. For now, let us quote here the simplest examples, which are given by}
\ba
\nonumber
A(1234)&\xrightarrow{s_{13}=0}
- \frac{1}{s_{12}} \sum_{\e_{\hat 3}} A(13-\!\hat 3)A(2\hat 34)\,,
\\
\label{n51}
A(12345)&\xrightarrow{s_{13}=s_{14}=0}
 \frac{1}{s_{12}} \big( 
\sum_{\e_{\hat 3}} A(13-\!\hat 3)A(2\hat 345) 
\nl
& \qquad\qquad+ \sum_{\e_{\hat 4}} A(14-\!\hat 4)A(23\hat 45) \big)\,.
\ea 

Moving on to the next case, $ m = 2 $, for amplitudes with $ n \geq 5 $ external legs we have
\ba 
&A({\mathbb I}_n)  \xrightarrow[s_{24}=s_{25}\ldots=s_{2\,n\!-\!1}=0]{s_{14}=s_{15}\ldots=s_{1\,n\!-\!1}=0}
\\ \nonumber 
& \quad (-1)^{n+1}\frac{s_{12}+s_{23}}{s_{12}s_{123}} \sum_{j=4}^{n-1}  \sum_{\e_{\hat j}} 
 A(1j\, -\!\hat j) A(234\ldots  \hat j   \ldots n)
 \\ \nonumber 
& \quad + (-1)^{n}\frac{s_{13}}{s_{12}s_{123}} \sum_{j=4}^{n-1}  \sum_{\e_{\hat j}} 
 A(2j\, -\!\hat j) A(134\ldots  \hat j   \ldots n)\,,
\ea 
where $ k_{\hat{j}} = k_1 + k_j $ and $ k_2 + k_j $ respectively in the last two lines implied by the momentum conservation of the three-point amplitudes in their lines.

Here we provide an example for $ n = 5 $,
\begin{align}
\label{n52}
A(12345) &\xrightarrow{s_{14} = s_{24} = 0}  \frac{s_{12} + s_{23}}{s_{12}s_{123}} \sum_{\e_{\hat 4}} A(14 -\!\hat{4}) A(23 \hat{4}5)
\nonumber
\\
& \quad - \frac{s_{13}} {s_{12}s_{123}} \sum_{\e_{\hat 4}}  A(24 -\!\hat{4}) A(13 \hat{4}5).
\end{align}
{In fact, it is easy to understand these new relations and its generalizations as a direct consequence of the BCJ identity. We will focus on the two examples \eqref{n51}  and \eqref{n52}. First applying the \textit{fundamental BCJ relation} \cite{Bern:2008qj} we have}
\begin{equation}
    A(12345)=\frac{s_{13}}{s_{12}}A(23145)+\frac{s_{13}+s_{14}}{s_{12}}A(23415)\,,
\end{equation}
{and we note that the $m=1$ limit $s_{13},s_{14}\to0$ induces a standard factorization in both terms of the RHS, i.e. $A_5 \to A_3 A_4$, thus verifying \eqref{n51}. In the general case, iteration of the fundamental BCJ relation reduces the amplitude to a $(n-3)!$-dimensional basis: An appropriate choice of this basis will be such that it has colinear singularities exactly where the zeros of $A(\mathbb{I})$ lie. For instance, for the case $m=2$ we iterate the fundamental relation twice to find}
\begin{equation}
    A(12345) = \frac{s_{12}+s_{23}}{s_{12}s_{123}} s_{14}\,A(23415) -\frac{s_{13}}{s_{12}s_{123}}s_{24}A(13425)\,,
\end{equation}
{thus proving \eqref{n52} via the same mechanism. Obviously, $ A(12345) \big|_{s_{13} = s_{14} = 0} $ and $ A(12345) \big|_{s_{14} = s_{24} = 0} $ are identified by relabeling, but a higher points the cases $m=1$ and $m=2$ are genuinely different. }
%
%
\subsection*{Soft Limit Interpretation }

{In the above formulae, momentum conservation in the three-point amplitude $A(ij {-}\hat{j})$  ensures that the internal gluon satisfies $k_{\hat{j}} = -k_{-\hat{j}} = k_i {+} k_j$. This behaviour is typical of soft limits and in fact provides an intuition of the above construction. Note that explicitly}
\ba 
\label{a1jcu}
A^\mu(ij;-\hat j) =    
\epsilon _i\!\cdot \!
   \epsilon _j\, k_i^\mu
   +\epsilon _i \!\cdot\! k_j \, \epsilon _j^\mu
   - 
   k_i\!\cdot \! \epsilon _j \,\epsilon _i^\mu\,,
   \ea 
{can be interpreted as an `exponential" soft factor including momentum and angular momentum terms \cite{inspirehep:2863391}. For instance, for the case $m=1$ we find}
\begin{align}\label{eq:exs}
    \sum_{j=4}^{n-1} \sum_{\e_{\hat j}} 
 A(1j\, -\!\hat j) A(234\ldots  \hat j   \ldots n) =&  \nonumber \\ 
 \sum_j\epsilon_1\cdot k_j \exp\left( \frac{F_1^{\mu \nu }  J^{(j)}_{\mu\nu}}{\epsilon_1 \cdot k_j}\right) & A(234   \ldots n)|_{\epsilon_1}\,,
\end{align}
where 
\begin{equation}
    J_{\mu\nu}= k_{[\mu} \frac{\partial}{\partial k^{\nu]}} +  \epsilon_{[\mu} \frac{\partial}{\partial\epsilon^{\nu]}}\,,
\end{equation}
{and we only keep terms linear in $\epsilon_1$. Note that each term is evaluated on the support of $k_1\cdot k_j$ which leads to the differential operator $\exp(k_1\cdot\partial_{k_j})$ generating the shift $k_{\hat{j}} = k_i+k_j$. }

{Finally, this allows us to interpret the new zeros of YM in a more physical way, namely as a type of \textbf{hard limit} (see e.g. \cite{Garcia-Sepulveda:2019tgx}). If we consider a regime where the planar propagators dominate $s_{12},s_{1n} \gg s_{ij} \gg s_{1i}$, where $i,j\neq n,1,2$ then the leading contribution emerges from the soft factorization \eqref{eq:exs}.} 

\subsection*{Main Claim}

Having provided the examples and intuition behind the new factorizations, we are ready to proceed with the precise general claim. We claim that the $n$-point YM amplitudes  with canonical ordering factorize into sums of gluings of a three-point and a $(n-1)$-point YM amplitude in the subspace of Mandelstam variables $ h_m = 0 $ for $m\geq 1$ and $ n \geq m+3 $,
\ba 
\label{totalcase}
&A({\mathbb I}_n)\! \xrightarrow{h_m=0}\!\! \!\!
\sum_{\substack{1\leq i \leq m <
m+2 \leq j \leq n-1}}
\!\!\!
F_{m,n}(i,j)\,,
\\
&F_{m,n}(i,j)=
\frac{(-1)^{n+1}}{s_{12\ldots m+1}} \sum_{\e_{\hat j}}
A(ij -\!\!\hat j)
\sum_{\rho \in S_{m-1}}  X(s,\rho)
\nonumber
\\
&\quad\times   A\big(\rho(12\ldots i\!-\!1\, i\!+\! 1 \ldots m)\, m{+}1\,m{+}2 \ldots { \hat j} \ldots n\big)\,.
\label{generalFn}
\ea

This factorization, illustrated in \cref{graph}, involves summing over all gluon pairs \((i, j)\) appearing in the matrix \( h_m \). Each term in the sum, denoted \( F_{m,n}(i,j) \), represents a ``gluon pair contribution''. Additionally, we sum over all possible polarization states of the internal gluon \( \hat{j} \), which is exchanged between the two amplitudes.

 Under the constraint \( h_m = 0 \), the Mandelstam variable \( s_{ij} \) vanishes, making the exchanged gluon on-shell. Finally, we sum over all permutations \( \rho \) of the \( m{-}1 \) gluons \( \{1,2, \dots, i{-}1, i{+}1, \dots, m\} \).

The terms \( X(s, \rho) \) are ratios of Mandelstam variables that involve only the gluons \( 1, 2, \ldots, m+1 \) and they remain independent of the total number of particles $n$. Examples of \( X(s, \rho) \) for low values of \( m \) are as follows,
\ba 
\nonumber
&\text{For } m=1, i=1, \quad  X(s, \emptyset) = 1 \,,\\
\nonumber
&\text{For } m=2, i=1, \quad X(s, 2) = \frac{s_{12}+s_{23}}{s_{12}}\,, \\
\label{examplexx}
&\text{For } m=2, i=2, \quad X(s, 1) =- \frac{s_{13}}{s_{12}} \,.
\ea 
In general, \( X(s, \rho) \) is expressed as
\ba 
\label{Xsrhoori}
  X(s, \rho) =& s_{12\ldots m{+}1}  \sum_{\pi \in \{ 12 \ldots i{-}1  \}\shuffle \{ i{+1} \ldots  m\} }   g^{-1}[\rho,\pi]  
  \nl
  & \times  \,  {\cal B}_{m+1,i}[   12 \ldots i{-}1   ,  i{+1} \ldots m | \pi ] \,,
\ea
where \( g^{-1} \) is the inverse of the matrix \( g[\rho, \pi] \), defined by
\ba 
\label{eq25g}
g[\rho,\pi]:= &\sum_{\substack{a=1\\ a\neq i}}^{m+1} 
 \left(\sum_{b\in Y(a,\rho)  } s_{ib} +s_{i\,m+1}\right)
 \nl &\times
  {\cal B}_{m+1,i}[Y(\rho, a) ,  Y(a,\rho) | \pi ] \,.
\ea 
Here, the shuffle operation \(  \shuffle \) combines two ordered sets and  in all possible ways while preserving the order of elements in each set. \( Y(\rho, a) \) represents the sequence of \( \rho \) that precedes \( a \), while \( Y(a, \rho) \) includes \( a \) and the part of \( \rho \) that follows it \footnote{For instance, if \( \rho = 12456 \) and \( a = 4 \), then \( Y(\rho, a) = 12 \) and \( Y(a, \rho) = 456 \).}. For \( a = m+1 \), we define \( Y(\rho, m+1) := \rho \) and \( Y(m+1, \rho) := \emptyset \). 

The term \( {\cal B}_{m+1,i}[\alpha, \beta | \pi] \) in \eqref{Xsrhoori} and \eqref{eq25g} is derived from the coefficients in the BCJ relations for YM amplitudes, as outlined in \cite{Bern:2019prr, Bern:2008qj},
\ba  
\label{BCJYMrelations}
A(\alpha\, i\, \beta \, m{+}1\, \hat n) = \sum_{\pi \in \alpha \shuffle \beta} A(i \, \pi\, m{+}1 \, \hat n) {\cal B}_{m+1,i}[\alpha, \beta | \pi].
\ea 
Here, \( k_{\hat n} \), determined by momentum conservation, is typically on-shell to maintain gauge invariance in YM amplitudes. However, in this context, the on-shell condition for \( k_{\hat n} \) is relaxed while ensuring that \( {\cal B}_{m+1,i}[\alpha, \beta | \pi] \) remains independent of \( k_{\hat n} \).

It turns out that the denominators of \( X(s, \rho) \) correspond to all planar poles involving gluon \( i \), ranging from \( 1 \) to \( m \). Numerically, this structure has been verified up to \( m=7 \). The numerators  are polynomials in the Mandelstam variables \( s_{rt} \) for \( 1 \leq r < t \leq m+1 \) and depend on the ordering \( \rho \).  Explicit forms of \( {X}(s, \rho) \) with polynomial numerators for \( m \leq 5 \) are provided in  ancillary files.

Finally, the contributions \( F_{m,n}(i, j_1) \) and \( F_{m,n}(m+1-i, j_2) \), where \( j_1, j_2 = m+2, m+3, \ldots, n-1 \) and \( j_1 + j_2 = n+m+1 \), are equivalent up to a sign under the relabeling \( a \leftrightarrow m+1-a \). As a result, \( X(s, \rho) \) satisfies the symmetry,
\ba
\label{symm}
X(s,\rho^T\big|_{a\to m+1-a \,{\rm mod} \, n })= X(s,\rho)\big|_{{a\to m+1-a \,{\rm mod} \, n }}\,,
\ea 
on the support of \( h_m = 0 \), i.e. $s_{rn} = -\sum_{t=1, t \neq r}^{m+1} s_{rt}$ for $1 \leq r \leq m$.

As an illustration, in \eqref{examplexx}, the term \( X(s, 1) \) can be rewritten as \( \frac{s_{12} + s_{1n}}{s_{12}} \) on the support of \( h_2 = 0 \), making the symmetry \eqref{symm} explicit when compared with the expression for \( X(s, 2) \).

A rigorous proof of the new factorization formula \eqref{totalcase} is provided in the companion paper \cite{Zhang:2024efe}.

\subsection*{Properties}

Applying the completeness relation of the polarization vector to our formula \eqref{generalFn}, we get a product of currents
\ba 
\label{completeness}
&\sum_{\e_{\hat j}}
\!
A(ij -\!\!\hat j) 
A\big(\rho  \ldots { \hat j} \ldots n\big) 
\!=\! A_\mu(ij ; -\hat j) A^\mu(\ldots n\, \rho \ldots ; \hat j ) ,
\ea 
with the three-point amplitude given by \eqref{a1jcu}.

Define $ x_{ij} := \{ \epsilon_i \cdot \epsilon_j, \epsilon_i \cdot k_j, k_i \cdot \epsilon_j\} $ and denote their collection as
\begin{align}
\label{eq:constrhatHm}
  \hat  H_m:=\begin{Bmatrix} 
    \hspace{-0.0cm} x_{1\,m+2}\, , \,x_{1\,m+3}\, , \ldots , x_{1\,n-1}\,\, , \\
    \hspace{-0.0cm} x_{2\,m+2}\, , \,x_{2\,m+3}\, , \ldots ,  x_{2\,n-1}\,\, , \\
    \hspace{-0.5cm}
    \vdots \hspace{1.1cm} \vdots \hspace{1.7cm} \vdots\\
    \hspace{-0.3cm}\hspace{0.3cm}x_{m\,m+2} , x_{m\,m+3} , \ldots , x_{m\,n-1}  ~~
    \end{Bmatrix}\,.
\end{align}
Additionally, denote $ H_m := h_m \sqcup \hat{H}_m $ where $\sqcup$ stands for the disjoint union . Setting $ x_{ij} = 0 $ obviously results in $ A^\mu(ij; -\hat{j}) = 0 $ as shown in \eqref{a1jcu}, hence $ F_{m,n}(i,j) = 0 $ following  \eqref{generalFn}. When setting all $ x_{ij} \in \hat{H}_m $ to zero, the entire amplitude $ A(\mathbb{I}_n) $ vanishes according to \eqref{totalcase},
\ba 
&A({\mathbb I}_n)\! \xrightarrow{H_m=0}
0\,,
\ea 
recovering the hidden zeros of the YM amplitudes as proposed in \cite{Arkani-Hamed:2023swr}. Our factorization formulas make this property manifest.

Conversely, when restoring a single $ x_{ij} \neq 0 $ from the constraint $ H_m = 0 $, the whole amplitude simplifies to a single gluon pair contribution,
\ba 
&A({\mathbb I}_n)\! \xrightarrow{(H_m/\{x_{ij}\})=0}
F_{m,n}(i,j)\,.
\ea 
This provides a physical meaning for the function $ F_{m,n}(i,j) $. Compared with the conventional factorization, our new factorization \eqref{totalcase} has an additional summation over the gluon pairs $(i, j)$. However, for each pair, we can assume that all other $ x_{i'j'} \neq x_{ij} \in \hat{H}_m $ vanish and consider its contribution individually. Then, their direct addition provides the whole factorization under the constraint $ h_m = 0 $.

When only one element, say $ \epsilon_i \cdot k_j \in x_{ij} $, is turned on instead of the whole $ x_{ij} $, only one term in the current $ A_\mu(ij; -\hat{j}) $ given in \eqref{a1jcu} survives, and the YM amplitude reduces to a combination of $(n-1)$-point amplitudes,
\ba 
A(\mathbb{I}_n) &\xrightarrow[{\rm except}~\epsilon_i \cdot k_j \neq 0]{H_m = 0} 
\frac{\epsilon_i \cdot k_j}{s_{12\ldots m+1}} 
\sum_{\rho \in S_{m-1}}  X(s,\rho)
\\
\nonumber
&\times   A\big(\rho(12\ldots i\!-\!1\, i\!+\! 1 \ldots m)\, m{+}1\,m{+}2 \ldots { \hat j} \ldots n\big)
\,,
\ea
where the momentum $ k_{\hat{j}} = k_i + k_j $ and polarization $ \epsilon_{\hat{j}} = \epsilon_j $ of the new gluon $\hat{j}$ are on-shell and orthogonal to each other on the support of the conditions $ H_m = 0 $ except $ \epsilon_i \cdot k_j \neq 0 $.  If only \( k_i \cdot \epsilon_j \in x_{ij} \) is turned on, \( \epsilon_{\hat{j}} = -\epsilon_i \). When turning on \( \epsilon_i \cdot \epsilon_j\)  instead, \( \epsilon_{\hat{j}} \) can take either \( k_i \) or \(-k_j\), reflecting the gauge redundancy of the amplitudes.

Following the discovery of zeros of YM amplitudes as described in \cite{Arkani-Hamed:2023swr}, a prompt inquiry has been the response of amplitudes near these zeros when the Lorentz products involving polarizations are turned back on.  Our factorization formulas provide a precise answer to this question.

In the following, we show more details for the cases $ m = 3, 4, 5 $ respectively.

\subsection*{Cases $ m = 3, 4, 5 $}

For $ m = 3 $, the gluon pair contributions    $ F_{3,n}(i,j) $ with $i=1,2,3$ are given by
\allowdisplaybreaks[1]
\ba 
\label{m3case1}
F_{3,n}(1,j)
&= 
\frac{(-1)^{n+1}}{s_{12}s_{123}s_{1234}} \sum_{\e_{\hat j}} A(1j -\!\hat j)
\nl
\times\Big(&
 -A(2345\ldots { \hat j} \ldots n) s_{2n} (s_{34}+s_{123})
\nonumber
\\
  &- A(3245\ldots { \hat j} \ldots n) s_{3n} s_{24}
\Big)
\,,
\\
\nonumber
F_{3,n}(2,j)
&= 
\frac{(-1)^{n+1}}{s_{12}s_{23}s_{123}s_{1234}} \sum_{\e_{\hat j}} A(2j -\!\hat j)
\\ 
\nonumber 
 \times
\Big(
&-A(1345\ldots { \hat j} \ldots n)\big(
s_{12} s_{14} s_{34}+s_{23}s_{3n} s_{1n}  
\\
&\qquad\qquad+s_{12} s_{23}(s_{13}+s_{14}+s_{3n})
\big)
\nonumber
\\
 &+ A(3145\ldots { \hat j} \ldots n) s_{14} s_{3n}(s_{12}+s_{23})\,
\Big)
\label{f3n2j}
\,,
\\ 
\label{m3case3}
F_{3,n}(3,j) &= 
\frac{(-1)^{n+1}}{s_{23}s_{123}s_{1234}} \sum_{\e_{\hat j}} A(3j -\!\hat j)
\nonumber
\\
 \times
\Big(
&-A(1245\ldots { \hat j} \ldots n) s_{24} (s_{1n}+s_{123})
\nonumber
\\
 &- A(2145\ldots { \hat j} \ldots n)s_{14} s_{2n}
\Big)\,.
\ea
\allowdisplaybreaks[0]
\!\!Summing over all such contributions gives the factorization of YM amplitudes under the condition \( h_3 = 0 \), as shown in \eqref{totalcase}. The explicit forms of \( X(s, \rho) \) for \( m = 3 \) can be directly derived from these contributions. For example, from the final term in \eqref{f3n2j}, we find \( X(s, 31) = \frac{s_{14}s_{3n}(s_{12}+s_{23})}{s_{12}s_{23}s_{123}} \) for \( F_{3,n}(2, j) \). Furthermore, it can be verified that \( X(s, 13) \) satisfies the symmetry property in \eqref{symm} under the relabeling \( a \leftrightarrow a' = n + 4 - a \, {\rm mod} \, n \) for all \( 1 \leq a \leq n \).

For $m = 4 $, each gluon pair contribution $ F_{4,n}(i,j) $ involves six $(n{-}1)$-point amplitudes, $ A(\rho(12\ldots i\!\!\slash  \ldots 4) 56 \ldots \hat{j} \ldots n) $ with $ \rho \in S_3 $. Their explicit expressions are present in the supplemental materials. All \( {X}(s, \rho) \) for \( m \leq 5 \) are also provided in ancillary files.

\section*{Further Factorizations}

Turning on a single $ x_{ij} $ from the constraint $ H_m = 0 $, the factorization of the YM amplitudes is described by a single gluon pair contribution $ F_{m,n}(i,j) $ given in \eqref{generalFn}. 
For $ m \geq 2 $, 
there are still many vanishing Lorentz products in the $(n-1)$ point amplitudes 
$A\big(\rho(12\ldots i\!-\!1\, i\!+\! 1 \ldots m)\, m{+}1\,m{+}2 \ldots { \hat j} \ldots n\big)$,
\ba 
s_{ab}&=\e_a\cdot k_b=k_a\cdot \e_b=\e_a\cdot \e_b=0,
\nl
\forall & a \in \{1,2,\ldots ,i\!-\!1, i\!+\! 1, \ldots ,m \},
\nl
& b\in \{m{+}2, m\!+ \! 3,\ldots j\!-\!1,j\!+\!1,\ldots, n\!-\!1\}\,.
\nonumber
\ea 

Thus, we consider further possible reductions of the amplitudes. For Lorentz products involving $ k_{\hat{j}} = k_i + k_j $, we have
\ba
&s_{a\hat j}=k_a\cdot k_{\hat j}=k_a\cdot  k_i \neq 0,\quad
\e_a\cdot k_{\hat j}=\e_a\cdot  k_i \neq 0\,.
\nonumber
\ea 
Although $ s_{aj} \in h_m = 0 $, $ s_{a\hat{j}} $ does not vanish, prompting the further factorization of the $(n-1)$-point amplitudes $ A\big(\rho \ldots \hat{j} \ldots n\big) $ into a product of two currents based on the 2-split construction proposed in \cite{Cao:2024gln, Cao:2024qpp}, with additional Lorentz products involving polarizations needing to be set to zero.

These Lorentz products include $ \epsilon_a \cdot \epsilon_{b'} $ and $ k_a \cdot \epsilon_{b'} $ with $ b' \in \{\hat{j}, m+1, n\} $. However, after summing over all states of $ \epsilon_{\hat{j}} $ according to \eqref{completeness}, it is $ \epsilon_j^\mu, k_j^\mu $, or $ \epsilon_i^\mu $ in $ A^\mu(ij; -\hat{j}) $ that contract with the current $ A^\mu(\rho \ldots {\hat{j}} \ldots n) $. Hence, $ \epsilon_j^\mu, k_j^\mu $, or $ \epsilon_i^\mu $ play the role of $ \epsilon_{\hat{j}} $ after the state sum. Correspondingly, we examine their contractions with $ \epsilon_a $ or $ k_a $ and get
\ba 
\nonumber
& k_a\cdot\e_j=   k_a\cdot k_j =0,  \quad k_a\cdot \e_i \neq 0 \,,
\nl
& 
 \e_a\cdot\e_j=   \e_a\cdot k_j =0, \quad \e_a\cdot \e_i \neq 0\,.
 \nonumber
\ea 
Therefore, based on the 2-split construction proposed in \cite{Cao:2024gln, Cao:2024qpp}, we just need to further impose the conditions
\ba 
\label{addcon}
&\e_a\cdot \e_{b'}=k_a\cdot \e_{b'} =0\,,
\\
\nonumber
 & \quad \forall a \in \{1,2,\ldots ,i\!-\!1, i\!+\! 1, \ldots ,m \}, \quad b'\in \{i,m{+}1, n \}\,,
\ea 
so that all contractions between $ \epsilon_j^\mu, k_j^\mu $, or $ \epsilon_i^\mu $ and $ A^\mu(\rho \ldots {\hat{j}} \ldots n) $ split into two currents. 

In this way, we achieve a three-split of the YM amplitude for $ 2 \leq m \leq n-3 $, where it decomposes into a sum of products of three currents,
\ba 
\label{threesplit}
A(\mathbb I_n)  &\xrightarrow[\eqref{addcon}]{(H_m/\{x_{ij}\})=0} 
\frac{(-1)^{n+1}}{s_{12\ldots m+1}} 
A^{\mu}(ij;-\hat j)
\Big[
\sum_{\rho \in S_{m-1}}  X(s,\rho)
\nonumber
\\
&\times   J^{{\rm YM}+\phi^3}(n^\phi \rho(1 2 \ldots i\!-\!1 \, i\!+\! 1 \ldots m)\, (m{+}1)^\phi\, ; \tilde{j}'{}^{\phi}) 
\Big]
\nonumber
\\
& \times J^{\rm YM}_{\mu}(j\!+\!1\, j\!+\!2\ldots n \,m\!+\!1\, m\!+ \! 2\, \ldots j\!- \!1; \tilde{j} ) \,.
\ea 
Here, $ J^{{\rm YM} + \phi^3}(n^\phi \ldots; \tilde{j}'{}^\phi) $ are single-trace currents for a mixed theory, dubbed as ${\rm YM} + \phi^3$, with two on-shell scalars $ n, m+1 $ and one off-shell scalar $ \tilde{j}'{}^\phi $. $ J^{\rm YM}(j+1 \ldots j-1; \tilde{j}) $ is a YM current with an off-shell leg $ \tilde{j} $. The momenta of the off-shell particles $ \tilde{j}'{}^\phi $ and $ \tilde{j} $ are constrained by momentum conservation in their currents. In practice, one can use lower-point amplitude formulas to compute these currents, ensuring that all momenta of the off-shell particles including those in the Mandelstam variables and $s_{m\!+\!1\,n}$ are eliminated on the support of momentum conservation in each current.   For more details on the currents, refer to \cite{Cao:2024gln, Cao:2024qpp,Naculich:2015zha}.

Here is an example of \eqref{threesplit},
\ba 
A(12345)  &\xrightarrow[
{\substack{
\e_2\cdot \e_1 = \e_2\cdot \e_3 = \e_2\cdot \e_5 =0
\\
k_2\cdot \e_1 = k_2\cdot \e_3 = k_2\cdot \e_5 =0
}}
]{s_{14}=s_{24}=0,x_{24}=0} 
\frac{s_{12}+s_{23}}{s_{12}s_{123}} 
\\ \nonumber &\times
  A^{\mu}(14;-\hat 4)
  J^{{\rm YM}+\phi^3}(5^\phi 23^\phi ; \tilde{4}'{}^{\phi}) 
 J^{\rm YM}_{\mu}(53 ; \tilde{4} ) \,,
\ea 
where $ J^{{\rm YM} + \phi^3}(5^\phi 23^\phi; \tilde{4}'{}^\phi) = -\epsilon_2 \cdot k_3 / s_{23} + \epsilon_2 \cdot k_5 / s_{25} $ and $ J^{\rm YM}_\mu(53; \tilde{4}) = \epsilon _5\!\cdot \!
   \epsilon _3\, k_5^\mu
   +\epsilon _5 \!\cdot\! k_3 \, \epsilon _3^\mu
   - 
   k_5\!\cdot \! \epsilon _3 \,\epsilon _5^\mu$.

\section*{Discussion}

In this paper, we investigated a novel factorization of YM amplitudes by setting specific non-planar two-point Mandelstam variables in a rectangular configuration to zero, which is the same constraint used to reveal zeros in some scalar theories as proposed in \cite{Arkani-Hamed:2023swr}. Our new factorization formulas involve summing over gluon pairs appearing in the vanishing Mandelstam variable matrix. By setting all Lorentz products involving polarizations associated with the vanishing Mandelstam variables to zero, we observed how the contribution from each gluon pair in our factorization summation cancels out one by one, ultimately revealing the hidden zeros of YM amplitudes. In cases where only the contribution from a single gluon pair survives, setting additional Lorentz products involving polarizations to zero reduces the amplitudes into a sum of products of three currents. This work introduces a novel type of factorization, distinct from conventional methods that rely on residues at poles, and opens the possibility of recursively constructing amplitudes using our new factorization formulas.

In the companion paper \cite{Zhang:2024efe}, we will provide proof of our new factorization formulas using the Cachazo-He-Yuan formalism \cite{Cachazo:2013hca}. Under the constraint of vanishing Mandelstam variables, one can prove that all solutions to the scattering equations \cite{Cachazo:2013gna,Cachazo:2013iea} become singular, making it evident why the amplitudes for YM theory (and many other theories such as GR, in addition to scalar theories) vanish when certain Lorentz products are set to zero. We will then demonstrate how these singular solutions independently contribute to yield new factorization formulas for YM theories when Lorentz products involving polarizations are turned back on.

Returning to our current work, it would be highly valuable to interpret the ratios of Mandelstam variables $X(s, \rho)$ in \eqref{threesplit} as amplitudes or currents associated with specific theories, or even to uncover a Feynman rule that generates them. Exploring the new factorization at the level of Feynman diagrams offers another promising direction. This perspective could shed light on how each summand in the factorization operates independently within different Feynman diagrams, providing deeper insights into the physical significance of the hidden zeros in YM amplitudes \cite{Rodina:2024yfc,Zhou:2024ddy,Huang:2025blb,Zhou:2025tvq,Feng:2025dci,Li:2025suo,Zhou:2025xly} \footnote{See recent progress at loop level in \cite{Backus:2025hpn}}.

While this paper focuses on YM amplitudes, it is equally valuable to investigate potential new factorizations for other theories, such as GR, and their extensions in string theories, both bosonic and supersymmetric \cite{green1988superstring,  Mafra:2011nw, He:2016iqi, He:2018pol}. By examining the behavior of their amplitudes around the zeros, we could derive new factorization formulas for these theories. Additionally, exploring their connections to double copy constructions \cite{Kawai:1985xq, Bern:2008qj} and soft theorems \cite{weinberg1965infrared} would be highly beneficial. Extending this study from tree-level to loop-level amplitudes is another promising direction. By identifying zeros in loop integrands, we can further explore their new factorization properties.  Further insights can be gained from studying the curve integral formalism for ${\rm Tr}(\phi^3)$ amplitudes, which provides a framework for deriving YM amplitudes at all loop levels via the so-called ``scalar scaffolding'' procedure \cite{Arkani-Hamed:2023swr, Arkani-Hamed:2023lbd, Arkani-Hamed:2023mvg, Arkani-Hamed:2023jry, Arkani-Hamed:2024nhp, Arkani-Hamed:2024yvu, Arkani-Hamed:2024fyd, De:2024wsy,Arkani-Hamed:2024tzl}.

\begin{acknowledgments}
We are deeply grateful to Freddy Cachazo for suggesting this study, providing invaluable guidance during its early stages, and offering continued support and encouragement throughout the project.
Our thanks also go to Qu Cao,  Jin Dong, Song He and Bruno Umbert for their insightful discussions and constructive feedback on our manuscript. 
The research of Y.Z.\ was supported in part by a grant from the Gluskin Sheff/Onex Freeman Dyson Chair in Theoretical Physics and by Perimeter Institute. Research at Perimeter Institute is supported in part by the Government of Canada through the Department of Innovation, Science and Economic Development Canada and by the Province of Ontario through the Ministry of Colleges and Universities. AG acknowledges the Roger Dashen membership at the IAS, and additional support from DOE
grant DE-SC0009988.

{{\bf Comments added}:
This updated arXiv version adds a new section and a co-author who contributed to the relation between zeros, soft factors, and BCJ structures \cite{guevara:zeros}. Several clarifying explanations have also been incorporated, while the main derivations and results remain unchanged.}

\end{acknowledgments}




 \bibliographystyle{apsrev4-1}
 \bibliography{Refs.bib}

\widetext
\newpage
\begin{center}
\textbf{\large Supplemental materials}
\end{center}
\begin{appendix}

\section{New factorization for $m=4$ \label{sec_reviewFac}}

For $ m=4 $, the gluon pair contributions $ F_{4,n}(i, j)$ with $i=1,2$ are given by
\ba 
\label{m4case1}
 F_{4,n}(1,\,&j)= 
\frac{ 1}{s_{12}s_{123}s_{1234}s_{12345}} \sum_{\e_{\hat j}} A(1j -\!\hat j) 
\\
\times
\Big(
 & A(23456\ldots  {\hat j} \ldots n)  s_{2n} (s_{12}\!-\!s_{3n})
   (s_{4n}\!-\!s_{123})
   \nonumber
   \\+
 &A(24356\ldots  {\hat j} \ldots n)   s_{35} s_{2n}
   s_{4n} 
   \nonumber
   \\+
 & A(32456\ldots  {\hat j} \ldots n) (s_{24}\!+\!s_{25}) s_{3n}
   (s_{4n}\!-\!s_{123})
   \nonumber
   \\+
  &A(34256\ldots  {\hat j} \ldots n)   s_{25} s_{3n}
   (s_{4n}\!-\!s_{123})
   \nonumber
   \\
   -& A(42356\ldots  {\hat j} \ldots n)
   \nonumber(s_{12}\!+\!s_{23}\!+\!s_{25}) s_{35}
   s_{4n}
  \\+
  & A(43256\ldots  {\hat j} \ldots n) 
s_{25} (s_{12}\!-\!s_{35})
   s_{4n}
\Big)
\nonumber
\,,
\\
\label{m4case2} 
 F_{4,n}(2,\,&j)= 
\frac{1 }{s_{12}s_{23}s_{123}s_{234}s_{1234}s_{12345}} \sum_{\e_{\hat j}} A(2j -\!\hat j)
\\
 \times \Big( &A(13456\ldots  {\hat j} \ldots n) \big((s_{13}\!\!+\!\!s_{14}\!+\!s_{15}) s_{23}
   s_{234} s_{3n}
   (s_{13}\!+\!s_{23}\!-\!s_{4n})
   \nl&\qquad\qquad
   \!-\!s_{12}^2((s_{13}
   s_{23}\!+\!s_{14} (s_{23}\!+\!s_{34}\!+\!s_{35}))
   s_{234}\!+\!s_{15} (s_{23}\!+\!s_{34}\!+\!s_{35})
   (s_{45}\!+\!s_{234})) 
   \nl&\qquad\qquad
   \!-\!s_{12}\big(2
   s_{23} s_{234} s_{13}^2\!+\! s_{13}(s_{15}
   ((s_{34}\!+\!s_{35})
   (s_{45}\!+\!s_{234})\!+\!s_{23} (s_{45}\!+\!2
   s_{234}))
   \nl&\qquad\qquad
   \!+\!s_{234} (s_{14} (3
   s_{23}\!+\!s_{34}\!+\!s_{35})\!+\!s_{23}
   (s_{23}\!+\!s_{234}\!+\!s_{345})))
   \nl&\qquad\qquad
   \!+\!(s_{23}\!+\!s_{34}\!+\!s_{35})
   ((s_{14}\!+\!s_{15})
   s_{234}^2\!+\!(s_{14}\!+\!s_{15})
   (s_{14}\!+\!s_{23}\!+\!s_{45}) s_{234}\!+\!s_{15} s_{23}
   s_{45})\big) \big)
   \nonumber
   \\
  - & A(14356\ldots {\hat j} \ldots n)   s_{35} (s_{12}
   (s_{14} s_{234}
   (\!-\!s_{4n}\!+\!s_{13}\!+\!s_{23})\!+\!s_{15} (s_{14}
   s_{234}\!+\!(s_{24}\!+\!s_{34}\!+\!s_{45})
   (s_{123}\!+\!s_{234})))
   \nl&\qquad\qquad
   \!+\!s_{23} s_{234}
   ((s_{13}\!+\!s_{15})
   (s_{14}\!+\!s_{24}\!+\!s_{34}\!+\!s_{45})\!-\!s_{14}
   s_{4n})\!+\!s_{14} s_{234} s_{12}^2)
      \nonumber
   \\
  + & A(31456\ldots {\hat j} \ldots n) 
   s_{3n} ((s_{14}\!+\!s_{15}) s_{234} (s_{23}
   (\!-\!s_{4n}\!+\!s_{13}\!+\!s_{23})\!+\!s_{12}^2\!+\!(s_{13}\!+\!s_{1
   4}\!+\!s_{23}\!+\!s_{45}) s_{12})
   \nl&\qquad\qquad
   \!+\!s_{12}
   (s_{14}\!+\!s_{15}) s_{234}^2\!+\!s_{12} s_{15} s_{45}
   s_{123})
      \nonumber
   \\
  +&  A(34156\ldots {\hat j} \ldots n) 
 s_{15} s_{3n} (s_{23}
   s_{234} (\!-\!s_{4n}\!+\!s_{13}\!+\!s_{23})\!+\!s_{12}
   (s_{123}\!+\!s_{234})
   (s_{23}\!-\!s_{4n}))
      \nonumber
   \\
  + &A(41356\ldots  {\hat j} \ldots n) 
s_{35}s_{4n}
   ((s_{13}\!+\!s_{15}) s_{23} s_{234}\!+\!s_{12}
   s_{15} (s_{123}\!+\!s_{234}))  
      \nonumber
   \\
    +& A(43156\ldots {\hat j} \ldots n) 
 s_{15}s_{4n} (s_{23} s_{35}
   s_{234}\!+\!s_{12} (s_{13}\!+\!s_{35})
   (s_{123}\!+\!s_{234}))  
   \Big)
\,.
\nonumber
\ea 
Performing a relabeling $ a \to a' = n + 5 - a $ modulo $ n $ for all $ 1 \leq a \leq n $ on \eqref{m4case1} and \eqref{m4case2}, we obtain $ F_{4,n}(4,j) $ and $ F_{4,n}( 3,j) $, respectively.  Summing over such functions provides the factorization of YM amplitudes under $h_4=0$ according to \eqref{totalcase}.

\end{appendix}

\end{document}